# Photonic and electronic state interactions in BaTiO₃ based Optical Microcavity

*Jitendra Nath Acharyya[1], R.B. Gangineni[2], D. Narayana Rao[3], and G. Vijaya Prakash[1, \*]*

[1]*Nanophotonics Lab, Department of Physics, Indian Institute of Technology Delhi, New Delhi 110016, India.*
[2]*Department of Physics, Pondicherry, University, Puducherry 605014, India*
[3]*School of Physics, University of Hyderabad, Hyderabad-500046, India*
*\*Email: prakash@physics.iitd.ac.in*

**Abstract:** The photonic modes mediated absorption dynamics at femtosecond time scales along with the control of spontaneous emission tunability are investigated all-dielectric optical microcavity having BaTiO₃ (BTO) as defect layer. The cavity-enhanced transient absorption reveals the dominant excited state absorption (ESA) of both photonic and electronic modes due to strong third-order optical nonlinearity influence. Photoluminescence of BTO is found to be guided and tuned by the photonic cavity mode. Such active photonic structures can be envisaged as a potential candidate in nonlinear optics and photonic device applications.

## 1. Introduction:

A photonic crystal is a novel optical media exhibiting robust applications from nano-sensing to wave-guiding [1,2]. The very first observation of the inhibition of spontaneous emission by 3D photonic structure [3] and the optical field confinement [4] led to a new tailor-made photonic structure appealing to vivid applications in nanophotonic and nonlinear optics [5-7]. The 1D photonic crystal (optical microcavity) is realized from a central defect layer sandwiched between two distributed Bragg reflectors. The optical field confinement and the control of spontaneous emission can be visualized through the interaction of the embedded material's electronic states and the photonic density of states of the photonic architecture. Because of the intriguing properties of the optical field modulation and the controlled optical field propagation, this novel photonic structure can be envisaged in optical filer, optical switches, optical limiters, etc. [8-10].

The realization of the light-matter interaction in such novel optical media is not straightforward. Complexity arises due to the interaction of photonic modes and the electronic states by incorporating an active photonic material, which has attracted researchers in the past few decades [6,11,12]. Barium titanate, BaTiO$_3$ (BTO) is one of the layered perovskite structures which exhibits special electro-optical and magnetic properties because of the TiO$_6$ octahedra, which supports the Mott type exciton within the large electronic bandgap of the structure [5, 13, 14]. The photoluminescence of BTO based optical structures were recently reported [5, 15, 16]. In contrast, the enhanced features of electron-photon interaction and their related effects on excited-state absorption and tunable photoluminescence have not been explored in an active one-dimensional layered photonic structure. The present work reports the cavity-influenced photoluminescence and the ultrafast optical nonlinearities using femtosecond optical pump-probe and single-beam Z-scan techniques. The optical microcavity was designed by two distributed Bragg reflectors of (SiO$_2$/TiO$_2$) with the central layer of BTO is fabricated via RF sputtering method [Fig. 1a]. The detailed fabrication procedures were given elsewhere [15,16]. The optical thicknesses of the constituent layers are found to be 73±5 nm, 55±5 nm, and 199±5 nm for TiO$_2$, SiO$_2$, and BaTiO$_3$, respectively [5].

## 2. Result and Discussions:

### 2.1. Optical properties of BaTiO$_3$ microcavity

The photonic band structure of the optical microcavity is visualized from the linear transmission and reflection spectral measurement. Fig. 1b shows the reflection spectrum of an optical microcavity, where the blue dashed trace represents the TMM simulation. The photonic bandgap is found to be between 1.85 eV to 2.7 eV. The incorporation of BaTiO$_3$ as a central cavity layer gives rise to the photonic cavity mode at 2.23 eV at the zero-degree angle of incidence.

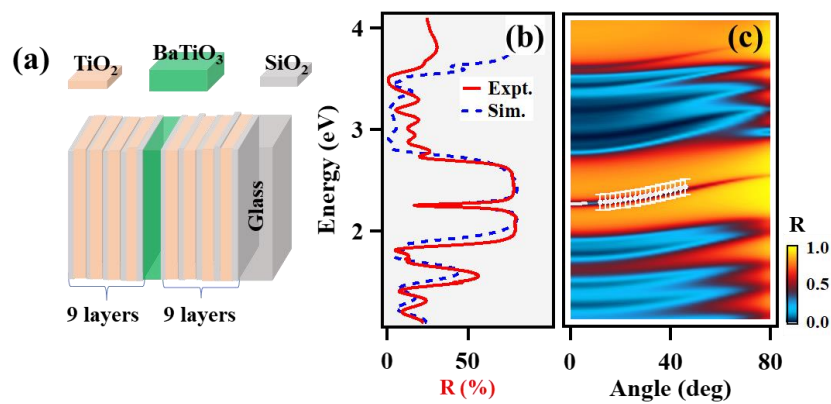

**Fig. 1.** (a) Schematic view of the BaTiO$_3$ microcavity. (b) experimental and transfer matrix simulated reflection spectra of the optical microcavity. (c) Angle-resolved reflection spectral map, wherein the white dots are experimental points.

In the angle-dependent reflection spectral map, depicted in Fig. 1c, the progressive blue shifting of photonic cavity mode can be attributed to the reduction of the photonic lattice parameter suffered by the exciting electromagnetic field [6], which can be expressed as follows

$$E_{ph}(\theta) = E_c (1 - (\sin^2\theta / n_{eff}^2))^{-1/2} \qquad (1)$$

Here $E_{ph}(\theta)$ is the angle-dependent photonic cavity mode and $E_c$ is the photonic cavity mode at a normal angle of incidence, and $n_{eff}$ is the effective refractive index of the optical microcavity.

As expected, the spontaneous emission of sandwiched BTO is hugely influenced by optical confinement due to central cavity photonic mode. The band-edge radiative recombination of BTO is completely inhibited by the photonic structure. The radiative recombination (photoluminescence) stems from the defect states of BTO guided by the photonic mode of the optical microcavity [17]. The angle-dependent PL study, photoexcited at 4.18 eV, reveals that the emission is guided by the transmission photonic cavity mode which is depicted in Fig. 2a. The shade spectrum is the PL of the bare BTO film.

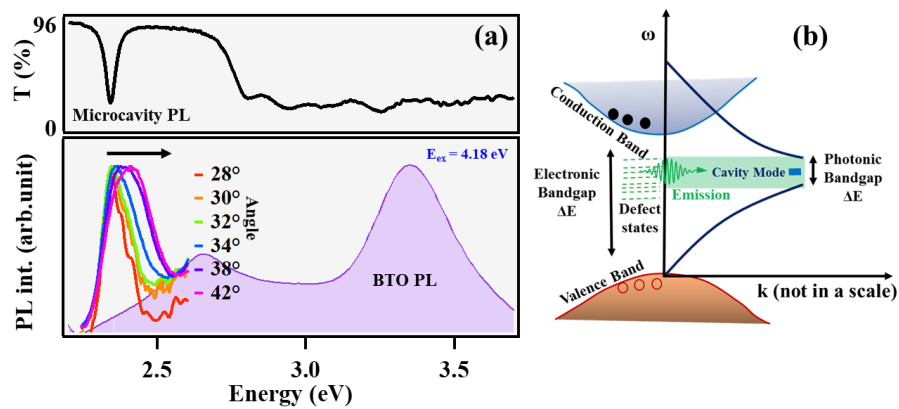

**Fig. 2. (a)** Angle dependent photoluminescence (PL) spectra of BaTiO$_3$ based optical microcavity. Shaded spectrum represents the PL for bare BaTiO$_3$ thin film (excited at $E_{ex}$ = 4.18 eV). Top graph is the reflection spectra at 30°. **(b)** Schematic representation of electronic and photonic energy band diagrams and the emission process in a photonic crystal.

## 2.2. Ultrafast dynamics and Nonlinear Optical response

The ultrafast dynamics are investigated using the femtosecond pump-probe technique using a pump beam centered at 3.54 eV (350 nm) of 120 fs pulse width at 1 kHz repetition rate and a broad-band white-light weak probe beam. Figure 3a represents the transient difference absorption ($\Delta A$) spectrum pumping with 0.25 µJ pulse energy. A positive $\Delta A$ (>0) peaks are observed for photonic cavity mode as well as for other photonic *miniband* regions. The positive $\Delta A$ (>0) is ascribed to the excited state absorption (ESA). As the pump beam excites, the defect energy states of $BaTiO_3$ are getting populated, which exhibits ESA in the presence of the probe beam. The observed time delays and probe energy $\Delta A$ contour map is depicted in Fig. 3b. The temporal dynamics of three distinct probed regions (indicated in Fig. 3a and top inset) are shown in Fig. 3c. The observed temporal dynamics show two distinct decay lifetimes, where the larger component is in the range of 15-50 ps and the shorter times in the range of 5-8 ps [5].

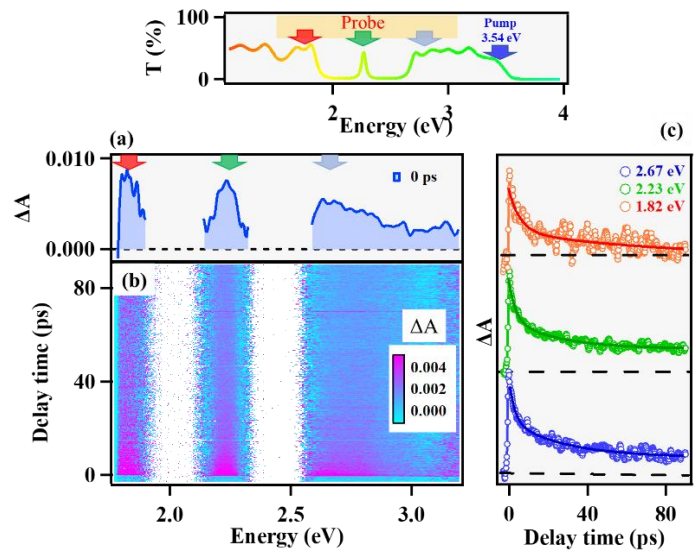

**Fig. 3.** Ultrafast absorption dynamics of the $BaTiO_3$ based optical microcavity performed by pumping at 3.54 eV (350 nm) laser beam (120 fs, 1 kHz, 0.25 µJ), and probed by broadband continuum beam. (a) Transient difference absorption ($\Delta A$) spectra at 0 ps. (b) $\Delta A$ spectral intensity map at different delay times. (c) Decay dynamics of $\Delta A$ signals probed at energies 1.82 eV, 2.23 eV, and 2.67 eV. (The top inset is the representation of the transmission spectrum with arrows indicating pump and probes regions).

For the same microcavity, the nonlinear optical absorption ($\beta$) and nonlinear refraction ($n_2$) were estimated using a single Gaussian beam Z-scan technique. In Z-scan, a laser beam centered at 1.55 eV (800 nm, 120 fs, 1 kHz) is focused with a convex lens, and the transmittance is recorded in two different modes of configuration. In one mode, an aperture is placed before the collection called closed aperture (CA) whereas, in open aperture (OA) configuration, all the transmitted is collected. The OA and CA scans are depicted in Figs. 4a and 4b, along with the respective theoretical fits [7,18-21]. The OA Z-scan trace (Fig. 4a) shows a rise in the transmittance at the focal position resulting due to the saturation absorption (SA) behavior which stems from the two-photon absorption (TPA) saturation [5]. The extracted TPA coefficient ($\beta$) is found to be 3.85 x 10$^{-5}$ m W$^{-1}$. The optical microcavity exhibits about seven orders enhancement in the TPA coefficient $\beta$ compared to the pure BaTiO$_3$ film [5]. The normalized CA scan (in Fig. 4b) exhibits a valley-peak configuration which signifies the positive nonlinear refraction behavior ($n_2 > 0$) originating from the self-focusing effect due to the electronic Kerr nonlinearity. The estimated $n_2$ is found to be ~ 6.97x10$^{-17}$ m$^2$ W$^{-1}$.

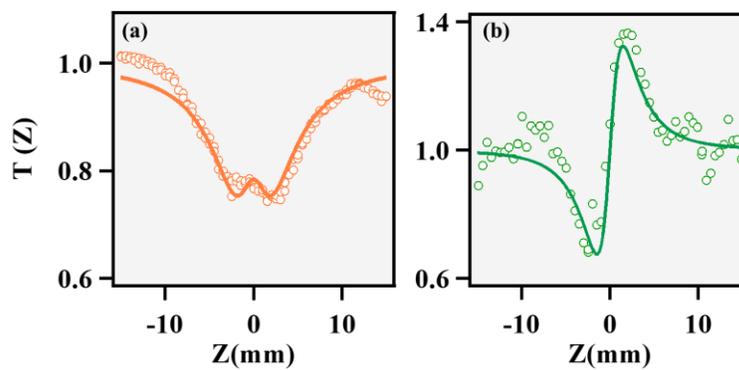

**Fig. 4.** The optical nonlinear was measured via single-beam Z-scan using femtosecond laser pulse centered at 1.55 eV (120 fs, 1 kHz). (a) Open (OA) and (b) Closed aperture (CA) Z-scan data of BaTiO$_3$ based optical microcavity. (Continuous curves are the theoretical fit).

## 3.   Conclusion

The electron-photon interactions are visualized through the BTO based one-dimensional (1D) photonic architecture. The linear optical properties agree well with the transfer matrix simulations. The cavity mode-assisted tunable photoluminescence behavior is strongly influenced by the photonic cavity. The transient absorption dynamics reveal the cavity influenced the excited state absorption behavior of BTO. The third-order nonlinear measurement of the optical microcavity reveals the enhancement in the two-photon absorption with respect to bare BTO film. The cavity-mediated photoluminescence established the control of materials' spontaneous emission by the active photonic microcavity. The novel 1D photonic architecture is a promising candidate for optical limiters, ultrafast laser operations, and future photonic device applications.

## Acknowledgments

Authors acknowledge SERB - DST (India) and Royal Society (UK) funding. The femtosecond laser facilities in UFO lab at IIT Delhi and the University of Hyderabad are acknowledged. JNA thanks DST-INSPIRE for the research fellowship.